\documentstyle[twocolumn,prb,aps,psfig,epsf,floats]{revtex}
\begin{document}
\title{Analytical and numerical study of hardcore bosons in two dimensions}
\author{K. Bernardet, G. G. Batrouni, J.-L. Meunier}
\address{
Institut Non-Lin\'eaire de Nice, Universit\'e de Nice--Sophia
Antipolis, 1361 route des Lucioles, 06560 Valbonne, France}
\author{G. Schmid, M. Troyer}
\address{Theoretische Physik, Eidgen\"ossische Technische Hochschule
Z\"urich, CH-8093 Z\"urich, Switzerland }
\author{A. Dorneich}
\address{Institut f\"ur Theoretische Physik, Universit\"at W\"urzburg,
97074 W\"urzburg, Germany}
\date{\today}
\address{\mbox{ }}        
%%%%%%%%%%%%%%%%%%%%%%%%%%%%%%%%%%%%%%%%%%%%%%%%%%%%%%%%%%%%%%%%%%%%
\address{\parbox{14cm}{\rm \mbox{ }\mbox{ }     
We study various properties of bosons in two dimensions interacting
only via onsite hardcore repulsion. In particular, we use the lattice
spin-wave approximation to calculate the ground state energy, the
density, the condensate density and the superfluid density in terms of
the chemical potential. We also calculate the excitation spectrum,
$\omega({\bf k})$. In addition, we performed high precision numerical
simulations using the stochastic series expansion algorithm. We find
that the spin-wave results describe extremely well the numerical
results over the {\it whole} density range $0\leq \rho \leq 1$. We
also compare the lattice spin-wave results with continuum results
obtained by summing the ladder diagrams at low density. We find that
for $\rho \leq 0.1$ there is good agreement, and that the difference
between the two methods vanishes as $\rho^2$ for $\rho \to 0$. This
offers the possibility of obtaining precise continuum results by
taking the continuum limit of the spin-wave results for all
densities. Finaly, we studied numerically the finite temperature phase
transition for the entire density range and compared with low density
predictions.}}
\address{\mbox{ }}
\address{\parbox{14cm}{\rm \mbox{ }\mbox{ }
}}
\address{\mbox{ }}
\maketitle

\section{Introduction}

The calculation of the thermodynamic properties of dilute Bose gases
is an old problem that keeps attracting renewed attention. Of
particular interest is the ground state energy per particle,
$\epsilon(\rho)$, where $\rho$ is the density of particles. The three
dimensional case, with hardcore repulsive interaction, was first
treated by Bogoliubov\cite{bog1}. This was later elaborated by many
other authors\cite{many} using a variety of methods. The main result
is that the leading term in the ground state energy per particle (in
three dimensions) is $\epsilon(\rho)=4\pi\rho a(\hbar^2/2m)$, where
$m$ is the mass of the particle and $a$ the scattering length. For
hardcore bosons, $a$ is simply the diameter of the particles. This
result was recently demonstrated rigorously by Lieb and
Yngvason.\cite{lieb1} The first two corrections to this result have
also been calculated.

Clearly, this result cannot be true in two dimensions since the units
would be incorrect. The units would work out if one eliminates the
dependence on the scattering length $a$, i.e. the hardcore boson
diameter. However, it is not reasonable to expect the energy to be
independent of the scattering length. The solution to this problem was
provided by Schick\cite{schick} in 1971 who used the field theoretic
methods developed and applied by Beliaev\cite{many} to the three
dimensional problem. This method, which applies in the very dilute
limit, consists of assuming that the dominant contributions come from
ladder diagrams which are then summed. Schick found that the leading
contribution to the energy per particle is $\epsilon(\rho)=4\pi
\rho(\hbar^2/2m) |{\rm ln}(\rho a^2)|^{-1}$ with the first correction
being proportional to $({\rm ln}(\rho a^2))^{-1}$. Hines {\it et al}
re-analyzed\cite{hines} the integral obtained by Schick and found that
the first correction term is larger and is equal to $ - {\rm ln}|{\rm
ln}({\bar \rho})|/|{\rm ln}({\bar
\rho})|$ where ${\bar \rho}= \rho a^2/\pi$. Lieb and
Yngvason\cite{lieb1} proved rigorously that the leading contribution
to the energy is indeed the term calculated by Schick, and that the
leading correction\cite{lieb2} is between ${\cal O}(|{\rm ln}(\rho
a^2)|^{-1})$ and ${\cal O}(|{\rm ln}(\rho a^2)|^{-1/5})$ which is
consistent with the correction calculated by Hines {\it et al}.

As mentioned above, these results are for very small densities.

In the meantime interest in hardcore bosons on two dimensional
lattices has been on the rise~\cite{fisherliu,ggb1,arovas,ggb2,ggb3}
and simulation algorithms have been improving dramatically.

In this paper we will study in detail hardcore bosons on a two
dimensional square lattice. Our approach is to do very high precision
numerical simulations using the stochastic series expansion (SSE)
algorithm\cite{sandvik,troyer} and mean field plus spin-wave analyses.
The quantities we calculate and measure with simulations are: The
density, $\rho$, as a function of the chemical potential, $\mu$, the
condensate, $\rho_0(\mu)$, the energy, $\epsilon(\mu)$, the excitation
spectrum and $\omega({\vec k})$. In addition, by twisting the boundary
conditions, the spin-wave calculation also yields the superfluid
density as a function of $\mu$ (or $\rho$). The numerical simulations
are done at very low temperature (essentially $T=0$) mostly on
$32\times 32$ systems with some $48\times 48$ results. We study these
systems from very low densities up to half filling, which is
equivalent to studying the entire density range $[0,1]$ due to the
particle-hole symmetry.

Our goal is to determine the range of densities for which the
spin-wave and continuum calculations are reliable. In addition we are
interested in whether and how the spin-wave calculation converges to
the continuum calculation (summation of ladder diagrams).

The paper is organized as follows. In section {\bf II} we present, in
some detail, the mean field and spin-wave calculations of the various
physical quantities at $T=0$. In section {\bf III} we compare the zero
temperature numerical results with the results of section {\bf II}. In
section {\bf IV} we compare the spin-wave calculation with the
continuum ladder diagram results valid for low densities. In section
{\bf V} we briefly examine finite temperatures and finally, in section
{\bf VI} we present some conclusions and comments.

\section{Mean field and spin-waves}

We will consider a two dimensional system of hardcore bosons on an
$L\times L$ lattice. The hamiltonian is given by

\begin{equation}
H = -t \sum_{\langle ij \rangle}(a^{\dagger}_i a_j + a^{\dagger}_j
a_i) - \mu \sum_i {\hat n_i}.
\label{hubham}
\end{equation}
where $\langle ij \rangle$ denotes near neighbors, $a_i$
($a^{\dagger}_i$) destroys (creates) a boson on site $i$, and $\mu$ is
the chemical potential. The hopping parameter, $t$, sets the energy
scale and will be set equal to one in the simulations. In addition, of
course, the hardcore constraint should be enforced. This means that
the number of bosons on a given site, $i$, as measured by the number
operator, ${\hat n_i}=a^{\dagger}_i a_i$, can only be $0$ or $1$. This
is not convenient to express in terms of the bosonic operators since
the normal commutation relations allow any number of bosons on a given
site.

To enforce the hardcore constraint in a simple way, we exploit the
exact mapping between this system and the two dimensional
antiferromagnetic Heisenberg model: $a_i^{\dagger} \leftrightarrow
S_i^{+}$, $a_i
\leftrightarrow S_i^{-}$, and ${\hat n_i}\leftrightarrow
S_i^z+1/2$. With this mapping the hardcore boson Hubbard hamiltonian
becomes

\begin{equation}
H = -t \sum_{\langle ij \rangle} (S_i^+ S_j^- + S_i^-S_j^+) -\mu
\sum_i S_i^z - {\mu \over 2} N,
\label{heisenham}
\end{equation}
where $N=L\times L$ is the total number of sites.

We start with the mean field calculation.

\subsection{Mean field}

The mean field calculation proceeds in the conventional
way\cite{ma}. The lowest possible energy state corresponds to all
spins down (no bosons): $|0\rangle = \prod_i |\downarrow\rangle_i$ and
a total magnetization of $-N/2$. Increasing $\mu$ to add bosons
corresponds to increasing the externally applied magnetic field in the
positive $z$ direction. In the mean field solution, we take all spins
parallel and making an angle $\theta$ with the positive $z$ axis. When
the system is empty $\theta=\pi$, and the (site independent) azimuthal
angle $\phi$ is $0$ for the moment. Increasing the chemical potential
(magnetic field) the spins will turn to reduce the angle with the $z$
axis. The mean field state vector is then

\begin{equation}
|\psi\rangle = \prod_i (u + v S_i^+)|0\rangle,
\label{MF}
\end{equation}
where $u={\rm sin}(\theta/2)$ is the amplitude for the spin to be
down, and $v={\rm cos}(\theta/2)$ the amplitude for an up spin. The
total $z$ spin is $S_z=S {\rm cos}(\theta)= N_b-N/2$ where $S$ is the
largest possible total spin ($=N/2$) and $N_b$ is the total number of
bosons. These equations yield the relation between the particle
density, $\rho=N_b/N$, and the angle $\theta$:

\begin{equation}
{\rm sin}^2(\theta) = 4\rho (1-\rho).
\label{rhoMF}
\end{equation}
Several quantities can now be readily calculated as functions of
$\rho$ or $\theta$. The expectation value of the hamiltonian in the
mean field state, $|\psi\rangle$, can be easily calculated and gives
the free energy per site, $F$ (recall that we are using the grand
canonical ensemble),

\begin{equation}
F = -t~{\rm sin}^2(\theta) -{{\mu}\over 2}{\rm cos}(\theta)- {\mu
\over 2}.
\label{F1}
\end{equation}
To determine the dependence of the chemical potential on the angle
$\theta$ we minimize $F$, $\partial F/\partial \theta =0$. This yields

\begin{equation}
{\rm cos}(\theta) = {\mu \over {4t}},
\label{thetamu}
\end{equation}
which, when used with Eq.~\ref{rhoMF}, gives the particle density in
terms of the chemical potential,

\begin{equation}
\rho = {1 \over 2} + {\mu \over {8t}}.
\label{rhomu}
\end{equation}
Therefore, $F$ becomes,

\begin{eqnarray}
F &=& -4t\rho^2, \nonumber \\
&=& -t\Bigl (1+ {\mu \over {4t}} \Bigr )^2.
\label{F2}
\end{eqnarray}

In the ground state, the internal energy (per site), $E$, which is
what is measured numerically, is related to the free energy via the
expression

\begin{eqnarray}
E &=& F + \mu \langle \rho \rangle, \nonumber \\
&=& -t + t \bigl ({\mu \over {4t}} \bigr )^2,\nonumber \\
&=& -4\rho t(1-\rho),
\label{E1}
\end{eqnarray}
where we used Eqs.~\ref{thetamu} and \ref{rhomu}.

The density of particles in the zero momentum mode (the condensate) is
given by:

\begin{eqnarray}
\rho_{0} &=& {1\over N} \langle \psi| {\tilde a}^{\dagger} ({\bf k}=0)
{\tilde a} ({\bf k}=0)|\psi \rangle,  \nonumber \\
&=& {1\over N} \sum_{i,j} \langle \psi|S_i^+ S_j^- |\psi\rangle,\nonumber \\
&=&  \rho (1-\rho),\nonumber \\
&=& {1\over 4}\Bigl ( 1- \bigl ({\mu \over {4t}}\bigr )^2\Bigr ),
\label{rho0MF}
\end{eqnarray}
where the operators ${\tilde a}^{\dagger}({\bf k})$ (${\tilde a}({\bf
k})$) are Fourier transforms of $a_i^{\dagger}$ ($a_i$). 

These results will be compared with the simulation results
below. First we calculate the spin-wave corrections.

\subsection{Spin-waves}

To include spin-wave corrections, we proceed the usual
way~\cite{fisherliu,cheng,ggb1,arovas}. We first rotate the $z$-axis
to align it with the mean field magnetization direction, {\it i.e.}
($\phi=0,\theta$). This is accomplished by the rotations

\begin{eqnarray}
S_i^x &=& S_i^{\prime x} {\rm cos}(\theta) + S_i^{\prime z} {\rm
sin}(\theta), \nonumber \\
S_i^y &=& S_i^{\prime y}, \\
S_i^z &=& -S_i^{\prime x}{\rm sin}(\theta) + S_i^{\prime z}{\rm
cos}(\theta),\nonumber
\label{rotate1}
\end{eqnarray}
where ${\vec S}_i^{\prime}$ is the spin vector in the rotated
frame. Equation~\ref{heisenham} becomes

\begin{eqnarray}
H &=& -2t \sum_{\langle ij\rangle} \Biggl ( \Bigl ( S_i^{\prime
x}S_j^{\prime x}{\rm cos}^2(\theta) + 2S_i^{\prime x}S_j^{\prime
z}{\rm cos}(\theta){\rm sin}(\theta) \nonumber \\
&&+ S_i^{\prime z}S_j^{\prime
z}{\rm sin}^2(\theta) \Bigr ) + S_i^{\prime y}S_j^{\prime y} \Biggr )
-\mu \sum_i \Bigl (-S_i^{\prime x} {\rm sin}(\theta)\nonumber \\
&& +S_i^{\prime z}
{\rm cos}(\theta) \Bigr ) - {\mu \over 2} N.
\label{heisenham2}
\end{eqnarray}
Now, the spins in the rotated frame are expressed in terms of bosonic
operators $b_i^{\dagger}$ and $b_i$. These operators do {\it not}
describe the original hardcore bosons, Eq.~\ref{hubham}. Instead they
describe the low energy bosonic fluctuations, the spin-waves. As such,
they obey the usual bosonic commutation relations,
$[b_i,b_j^{\dagger}]=\delta_{i,j}$. This transformation is
accomplished by using

\begin{eqnarray}
S_i^{\prime x}&=& {1\over 2}(S_i^{\prime +}+S_i^{\prime -}) = {1\over
2}(b_i^{\dagger}+b_i),\nonumber \\ 
S_i^{\prime y}&=& {1\over 2i}(S_i^{\prime +}-S_i^{\prime -}) =
{1\over 2{\rm i}}(b_i^{\dagger}-b_i), \\
S_i^{\prime z}&=& {1\over 2} - b_i^{\dagger}b_i.\nonumber
\label{bbosons}
\end{eqnarray}
Substituting these expressions in Eq.~\ref{heisenham2}, we assume a
dilute gas of spin-waves and ignore cubic and quartic terms in the
bosonic operators. This yields,

\begin{eqnarray}
H &=& -{t\over 2}\sum_{\langle ij\rangle}\Biggl (-{\rm
sin}^2(\theta) (b_i^{\dagger}b_j^{\dagger}+b_i b_j) \nonumber \\
&&+ (1+{\rm
cos}^2(\theta)) (b_i^{\dagger}b_j+b_i b_j^{\dagger}) \Biggr )
\nonumber \\
&&+ (4t~{\rm sin}^2(\theta) + \mu~ {\rm cos} (\theta)) \sum_i
b_i^{\dagger}b_i\nonumber \\ &&-N\Bigl ({\mu \over 2} (1+{\rm cos}
(\theta))+t~{\rm sin}^2(\theta)\Bigr )\nonumber \\ &&+\bigl (-2t~{\rm
cos} (\theta)+{\mu \over 2}\bigr ) {\rm sin} (\theta)
\sum_i(b_i^{\dagger}+b_i).
\label{spinham}
\end{eqnarray}
The term linear in the operators is required to vanish yielding,
\begin{equation}
{\rm cos}(\theta) = {\mu \over {4t}}.
\label{thetamu2}
\end{equation}
This is the same relation between $\theta$ and $\mu$ we found in the
mean field case: Spin-waves do not modify it. 

We are, therefore, left with a quadratic hamiltonian which we can
diagonalize by first going to Fourier space,

\begin{eqnarray}
b_{\bf r}^{\dagger} &=& {1\over{\sqrt{N}}} \sum_{\bf k} e^{-i{\bf
k.r}} b^{\dagger}_{\bf k},\nonumber \\ b_{\bf r} &=&
{1\over{\sqrt{N}}} \sum_{\bf k} e^{i{\bf k.r}} b_{\bf k}.
\label{FT}
\end{eqnarray}
This give,

\begin{eqnarray}
H &=& H_{MF} + \sum_{\bf k} \Bigl ( A_k \bigl (b^{\dagger}_{\bf k}b_{\bf
k}+b^{\dagger}_{\bf -k}b_{\bf -k}\bigr )  \nonumber \\
&& +B_k \bigl (b^{\dagger}_{\bf
k}b^{\dagger}_{\bf -k}+b_{\bf k}b_{\bf -k}\bigr ) \Bigr ),
\label{Hsw}
\end{eqnarray}
where,

\begin{eqnarray}
A_k &=& -{t\over 2}\Bigl ( \bigl (1+({\mu \over {4t}})^2\bigr
)\gamma_{\bf k} -4 \Bigr ),\nonumber \\
B_k &=& {t\over 2} \bigl (1-({\mu \over {4t}})^2\bigr )\gamma_{\bf
k},\nonumber \\
\gamma_{\bf k} &=& {\rm cos}(k_x)+{\rm cos}(k_y).
\label{ABG}
\end{eqnarray}
$H_{MF}$ is the constant term in Eq.~\ref{spinham}, which is
identical to Eq.~\ref{F1}. Eliminating $\theta$ by using
Eq.~\ref{thetamu2} we get the second of Eq.~\ref{F2}. In other
words, $H_{MF}$ is just the pure mean field result.

To diagonalize the quadratic spin-wave hamiltonian, Eq.~\ref{Hsw}, we 
use the Bogoliubov transformation,

\begin{equation}
b_{\bf k} = u_k \alpha_{\bf k}-v_k \alpha_{\bf -k}^{\dagger},~~~~~~
b_{\bf k}^{\dagger}= u_k \alpha_{\bf k}^{\dagger}-v_k \alpha_{\bf -k},
\label{bogo}
\end{equation}
where $\alpha_{\bf k}^{\dagger}$ ($\alpha_{\bf k}$) is a creation
(destruction) operator for a quasi-particle excitation of momentum
${\bf k}$. These operators satisfy the bosonic commutation relation,
$[\alpha_{\bf k},\alpha_{\bf k^{\prime}}^{\dagger}]=\delta_{\bf
k,k^{\prime}}$, if $u_k^2-v_k^2=1$. We can ensure this by writing,

\begin{equation}
u_k = {\rm cosh}(\phi_k),~~~~v_k = {\rm sinh}(\phi_k),
\label{phis}
\end{equation}
where $\phi_k$ is determined by the requirement the Bogoliubov
transformation diagonalize the hamiltonian, Eq.~\ref{Hsw}.

We, therefore, obtain

\begin{eqnarray}
{\rm sinh}^2(\phi_k) &=& {1\over 2} \Bigl ( {{A_k}\over
{\sqrt{A_k^2-B_k^2}}} - {1\over 2} \Bigr ),\nonumber \\
{\rm cosh}^2(\phi_k) &=& {1\over 2} \Bigl ( {{A_k}\over
{\sqrt{A_k^2-B_k^2}}} + {1\over 2} \Bigr ),
\label{sinhcosh}
\end{eqnarray}
with $A_k$ and $B_k$ given by Eq.~\ref{ABG}. The diagonalized
spin-wave hamiltonian then becomes,

\begin{eqnarray}
H &=& \Bigl ( H_{MF} + \sum_{\bf k} \bigl (\sqrt{A_k^2-B_k^2}-A_k\bigr
) \Bigr )\nonumber \\
&& + \sum_{\bf k\ne 0} \sqrt{A_k^2-B_k^2}\bigl (\alpha_{\bf
k}^{\dagger}\alpha_{\bf k} + \alpha_{\bf -k}^{\dagger}\alpha_{\bf -k}
\bigr ).
\label{HswDiag}
\end{eqnarray}
This immediately gives the spin-wave corrected free energy per site,
$F_{SW}$,

\begin{equation}
F_{SW} = -t\Bigl (1+ \bigl ({\mu \over {4t}}\bigr ) \Bigr )^2 +
{1\over N}\sum_{\bf k} \bigl (\sqrt{A_k^2-B_k^2}-A_k\bigr ),
\label{FSW}
\end{equation} 
where we wrote explicitly the expression for $H_{MF}$ using
Eq.~\ref{F2}. Equation~\ref{HswDiag} also gives the dispersion of
quasi-particle excitations,

\begin{equation}
\omega({\bf k}) = 2\sqrt{A_k^2-B_k^2}.
\label{omega}
\end{equation}
It is easy to show, using Eqs.~(\ref{ABG},\ref{omega}), that as ${\bf
k}\to 0$, we obtain the linear dispersion characteristic of phonons,

\begin{equation}
\omega({\bf k}\to 0) = 2t\sqrt{1-\bigl ({\mu \over {4t}}\bigr )^2}
|{\bf k}|.
\label{phonon}
\end{equation}
This gives the sound speed (critical velocity) as

\begin{equation}
c = 2t\sqrt{1-\bigl ({\mu \over {4t}}\bigr )^2}.
\label{sound}
\end{equation}

Now that we have diagonalized the hamiltonian, we can calculate the
spin-wave corrections to the quantities already obtained by mean
field. The number of particles kicked out of the condensate (${\bf
k}=0$ mode) due to the spin-waves is given by $n_{\bf k}=\langle
\psi|b^{\dagger}_{\bf k}b_{\bf k}|\psi\rangle $. Using
Eq.~\ref{bogo} and the fact that $\alpha_{\bf k}|\psi\rangle = 0$
(no quasi-particle excitations in the ground state), we get,

\begin{equation}
n_{\bf k\ne 0} = {1\over 2} \Bigl ( {{A_k}\over {\sqrt{A_k^2-B_k^2}}} -
{1\over 2} \Bigr ).
\label{nk}
\end{equation}
The spin-wave corrected condensate density is then,

\begin{equation}
\rho_0 = \rho\bigl (1-\rho \bigr ) - {1\over N}\sum_{\bf k \ne 0}
n_{\bf k}.
\label{rho0SW}
\end{equation}

Whereas spin-waves did not modify the relation between $\mu$ and
$\theta$, they do introduce a correction to the relation between $\mu$
and $\rho$. To find the new expression, we simply use $\rho =
-\partial F_{SW}/\partial \mu$. This gives

\begin{equation}
\rho = {1\over 2}\bigl (1+{\mu \over {4t}}\bigr ) + {1\over N}{\mu
\over {16t}}\sum_{\bf k\ne 0} \gamma_k \Biggl ({{A_k-B_k}\over
{\sqrt{A_k^2-B_k^2}}}-1\Biggr ).
\label{rhoSW}
\end{equation}

Now that we have determined $F_{SW}$ and $\rho(\mu)$, we can calculate
the boson energy per site (internal energy), $E$, using
$E_{SW}=F_{SW}+\mu\langle \rho \rangle$. This gives,

\begin{eqnarray}
E_{SW} &=& -t\Bigl (1- \bigl ({\mu \over {4t}}\bigr )^2 \Bigr ) +
{1\over N} \sum_{\bf k\ne 0} \bigl (\sqrt{A_k^2-B_k^2}-A_k\bigr
)\nonumber \\ 
&&+ {t\over N}\bigl ({{\mu} \over {4t}}\bigr )^2\sum_{\bf k\ne 0}
\gamma_k \Biggl ({{A_k-B_k}\over {\sqrt{A_k^2-B_k^2}}}-1\Biggr ).
\label{ESW}
\end{eqnarray}

\subsection{Superfluid density}

Unlike the above thermodynamic quantities, the superfluid density,
$\rho_s$, requires special treatment of the boundary conditions. As is
well known~\cite{MFisher}, one can relate $\rho_s$ to the ``spin
stiffness''. To accomplish this, we need to compare the free energy of 
the system under periodic and twisted boundary conditions. This energy 
difference is related to $\rho_s$.

In the periodic case, which we treated above, the azimuthal angle,
$\phi$, was taken to be site independent (equal to zero). To implement
a twist, we take this angle to be site dependent, $\phi_{\bf r}$, and
with a constant gradient such that the total twist across the system
in both the $x$ and $y$ directions is $2\pi$,

\begin{equation}
\delta \phi = \phi_{\bf r+{\hat x}}-\phi_{\bf r}=\phi_{\bf r+{\hat
y}}-\phi_{\bf r} = {{2\pi}\over L},
\label{deltaphi}
\end{equation}
where ${\bf \hat x}$ (${\bf \hat y}$) is a unit vector in the $x$ ($y$)
direction. Consequently, the rotation of the reference frame to align
it with the local spin will be site dependent,

\begin{eqnarray}
S_i^x &=& \Bigl (S_i^{\prime x} {\rm cos}(\theta) + S_i^{\prime z}{\rm
sin}(\theta)\Bigr ){\rm cos}(\phi_i)-S_i^{\prime y}{\rm sin}(\phi_i),
\nonumber \\
S_i^y &=& \Bigl (S_i^{\prime x} {\rm cos}(\theta) + S_i^{\prime z}
{\rm sin}(\theta)\Bigr ){\rm sin}(\phi_i)+S_i^{\prime y}{\rm
cos}(\phi_i), \\
S_i^z &=& -S_i^{\prime x}{\rm sin}(\theta) + S_i^{\prime z}{\rm
cos}(\theta).\nonumber
\label{rotate2}
\end{eqnarray}

The rotation and diagonalization now proceed exactly as before
giving,

\begin{equation}
\mu = 4t~{\rm cos}(\theta) {\rm cos}(\delta \phi),
\label{thetamutwist}
\end{equation}
and

\begin{eqnarray}
F^T_{SW} &=& -t\Bigl (1+ \bigl ({\mu \over {4t~{\rm cos}(\delta
\phi)}}\bigr ) \Bigr )^2 \nonumber \\
&&+ {1\over N}\sum_{\bf k} \bigl
( \sqrt{(A_k^T)^2-(B_k^T)^2}-A^T_k\bigr ),
\label{FSWT}
\end{eqnarray} 
where the $T$ superscript indicates that the boundary conditions are
twisted. $A^T_k$ and $B_k^T$ are given by

\begin{eqnarray}
A^T_k &=& -{t\over 2}\Biggl ( \Bigl (1+\bigl ({\mu \over {4t~{\rm
cos}(\delta \phi)}}\bigr )^2\Bigr )\gamma_{\bf k} -4 \Biggr
),\nonumber \\ 
B^T_k &=& {t\over 2} \Bigl (1-\bigl ({\mu \over {4t~{\rm
cos}(\delta \phi)}}\bigr )^2\Bigr )\gamma_{\bf k},
\label{ABT}
\end{eqnarray}
and $\gamma_{\bf k}$ has not changed.

We can now calculate $\Delta F_{SW} = F^T_{SW}-F_{SW}$, 

\begin{eqnarray}
\Delta F_{SW} &=& \Biggl ( \rho \bigl (1-\rho \bigr )\nonumber \\
&&+{1\over {tN}}\sum_{\bf k\ne 0}\Bigl (A_k -
\sqrt{A_k^2-B_k^2}\Bigr ) \Biggr ) t{{(\delta \phi)^2}\over 2},
\label{deltaF}
\end{eqnarray}
where we made the approximation ${\rm cos}(\delta \phi)\approx
1-(\delta \phi)^2/2$. The square of the gradient twist can be related
to the kinetic energy of the superfluid~\cite{MFisher} by 

\begin{equation}
(\delta \phi)^2= {{m^2}\over {2\hbar^2}} v^2,
\label{KE}
\end{equation}
with $v^2=v^2_x+v^2_y$. Recalling that $t=\hbar^2/(2m)$, we obtain the 
superfluid density,

\begin{equation}
\rho_s =  \rho \bigl (1-\rho \bigr )
+{1\over {tN}}\sum_{\bf k\ne 0}\Bigl (A_k -
\sqrt{A_k^2-B_k^2}\Bigr ).
\label{rhos}
\end{equation}

It is very interesting to note that at the mean field level, the
superfluid density and the condensate are identical, $\rho_0=\rho_s$
(see Eqs.(\ref{rho0SW},\ref{rhos})). As expected, spin-wave
corrections deplete the condensate, Eq.~\ref{rho0SW}, but
surprisingly, the superfluid density is {\it enhanced} since the
second term in Eq.~\ref{rhos} is positive. Furthermore, both our
spin-wave calculation and numerical simulations (see below) show that
even at zero temperature (the present case), the superfluid density is
never truly equal to the density nore to the condensate density. In
fact, we always have $\rho_s < \rho$ and, of course, $\rho_s >
\rho_0$. It is worth noting that at the very small densities present
in atomic condensates (of the order of $10^{-4}$), our results do
indeed yield $\rho_s\approx\rho_0\approx\rho$, which agrees with the
experimental observations.~\cite{kaiser}

We now compare these results with numerical simulations.

\section{Numerical simulations}

The numerical simulations were done using the SSE
algorithm.~\cite{sandvik,troyer} The results we will present here are
for a $32\times 32$ system with periodic boundary conditions and
inverse temperature $\beta=45$. Since we are, mostly, interested in
the zero temperature properties, we checked the finite temperature
effects by doing simulations at $\beta=60$. In addition, we checked
the finite size effects by simulating $48\times 48$ systems. Within
our very small error bars, the results were identical to the
simulations for $32\times 32$ at $\beta=45$. We can, therefore,
consider these results to be valid in the thermodynamic limit at
$T=0$.

One of the quantities of prime interest is the superfluid density,
$\rho_s$. In the previous section, we calculated $\rho_s$ using the
relation between the helicity modulus and the difference in free
energies with periodic and antiperiodic boundary
conditions.\cite{MFisher} Numerically, $\rho_s$ is determined from the 
winding number,~\cite{roy}

\begin{equation}
\rho_s =  {m \over {\hbar^2}}{{\langle W^2 \rangle} \over {D \beta}},
\label{rhosW1}
\end{equation}
where $W$ is the winding number, $D$ the dimensionality, and
$\beta=1/k_B T$. Since the hopping parameter $t$ is related to the
mass via $t=\hbar^2/(2m a^2)$, where $a$ is the diameter of the
hardcore boson (one lattice spacing in our case), Eq.~\ref{rhosW1}
becomes 

\begin{equation}
\rho_s =  {1\over {2ta^2}} {{\langle W^2 \rangle} \over {D \beta}}.
\label{rhosW2}
\end{equation}

In section {\bf V} we will consider the finite temperature KT
transition from superfluid to normal phases. This temperature from the 
superfluid density using the universal jump
condition\cite{Kosterlitz},

\begin{equation}
{\pi \over 2} \rho_s =  {{mk_B T_{KT}} \over \hbar^2}, 
\label{rhosKT}
\end{equation}
with $m$ related to $t$ as above.

The spin-wave results contain momentum summations which we did
numerically, typically for $L=32$, to compare directly with the
numerical simulations.

\begin{figure}
\psfig{file=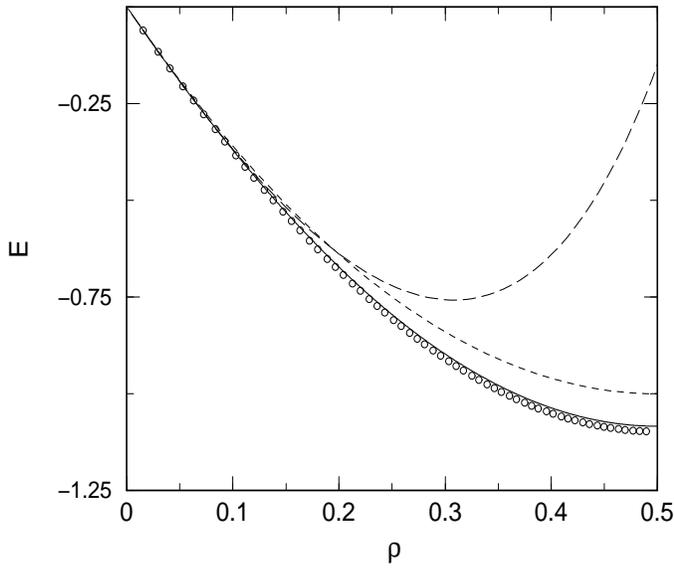,height=3.0in,width=3.5in}
\caption{The boson energy, $E$, vs. the density, $\rho$. Circles:
Simulation results, dashed line: Mean field, solid line: with
spin-wave correction, long dashed: Ladder diagram summation of Hines
{\it et. al.}\cite{hines} }
\label{Evsrho-L32}
\end{figure}

In Fig.~\ref{Evsrho-L32} we show the bosonic energy, $E$ as a function
of the particle density, $\rho$. The circles are the numerical
results, with error bars smaller than the size of the points. The
dashed line shows the mean field result, Eq.~\ref{E1}, and the solid
line shows the mean field plus spin-wave correction,
Eq.~\ref{ESW}. The long dashed line will be discussed in the next
section. We see that agreement between the numerical and spin-wave
results is excellent for the whole range of densities, $0\leq \rho
\leq 0.5$, and particle hole symmetry extends this agreement all the
way to full filling.

\begin{figure}
\psfig{file=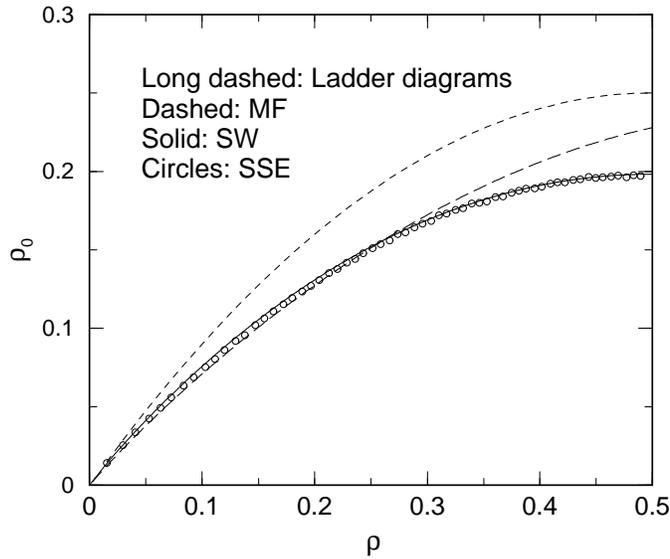,height=3.0in,width=3.5in}
\caption{The condensate fraction, $\rho_0$, vs. the density,
$\rho$. Circles: Simulation results, dashed line: Mean field, solid
line: with spin-wave correction, long dashed: Ladder diagram summation
of Hines {\it et. al.}\cite{hines} }
\label{R0vsrho-L32}
\end{figure}

Figure~\ref{R0vsrho-L32} shows the comparison of the condensate
fraction, $\rho_0$, Eqs.~\ref{rho0MF} and ~\ref{rho0SW} with the
numerical results. The line convention is as for the previous
figure. Once again we see that the agreement between spin-waves and
simulations is remarkable over the entire density range. Numerically, the
condensate fraction $\rho_0$ was calculated as the Fourier transformed of the
zero momentum equal time Greens function.

\begin{figure}
\psfig{file=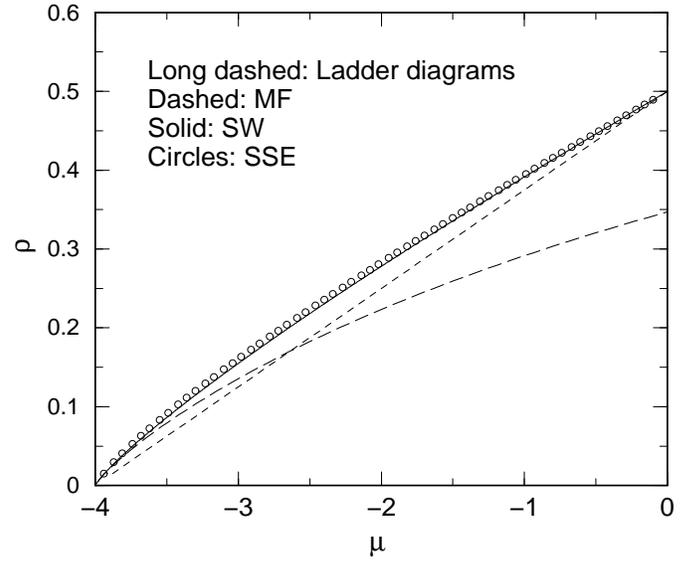,height=3.0in,width=3.5in}
\caption{The particle density, $\rho$, vs. the chemical potential,
$\mu$. Circles: Simulation results, dashed line: Mean field, solid
line: with spin-wave correction, long dashed: Ladder diagram summation
of Hines {\it et. al.}\cite{hines} }
\label{rhovsmu-L32}
\end{figure}

In Fig.~\ref{rhovsmu-L32} we show the dependence of the particle
density on the chemical potential, Eqs.~\ref{rhomu} and
~\ref{rhoSW}. Again we see that the agreement between the spin-wave
calculation and the numerical simulation is excellent.

\begin{figure}
\psfig{file=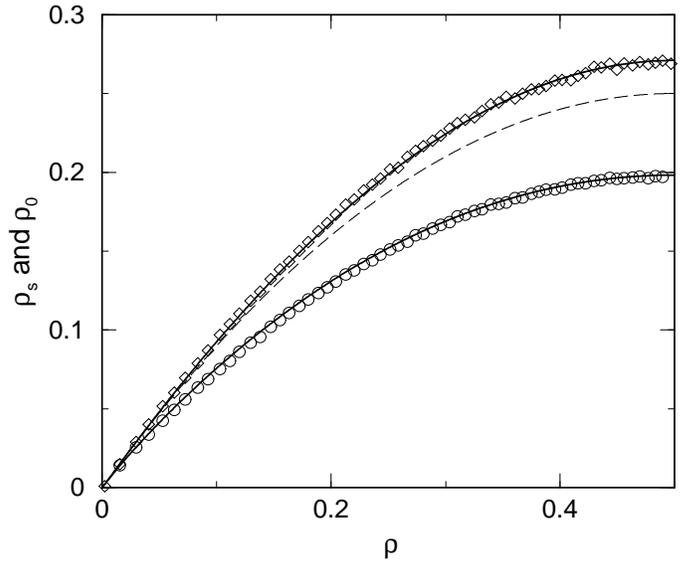,height=3.0in,width=3.5in}
\caption{The dashed line gives both, $\rho_0$ and $\rho_s$
vs. $\rho$. The numerical results are shown in diamonds, $\rho_s$, and
circles, $\rho_0$. The solid lines are the corresponding spin-wave
results.}
\label{RsR0vsrho}
\end{figure}

The superfluid density, $\rho_s$ (Eq.~\ref{rhos}) is shown with the
condensate fraction, $\rho_0$ (Eq.~\ref{rhoSW}), as functions of
$\rho$ in Fig.~\ref{RsR0vsrho}. The dashed line is the mean field
result for both quantities. At the mean field level, $\rho_0$ and
$\rho_s$ are identical which is, perhaps, not so surprising. What is
surprising is that when we include the spin-waves, $\rho_0$ is pushed
down while $\rho_s$ is enhanced. As we see from the figure, agreement
with numerical results is extremely good. 

It is clear from Fig.~\ref{RsR0vsrho} that for small densities we have
$\rho_s \approx \rho$ and $\rho_0 \approx \rho$. This holds well for
$\rho \leq 0.03$ since at this density $\rho_s$ and $\rho_0$ are no
longer equal.

\begin{figure}
\psfig{file=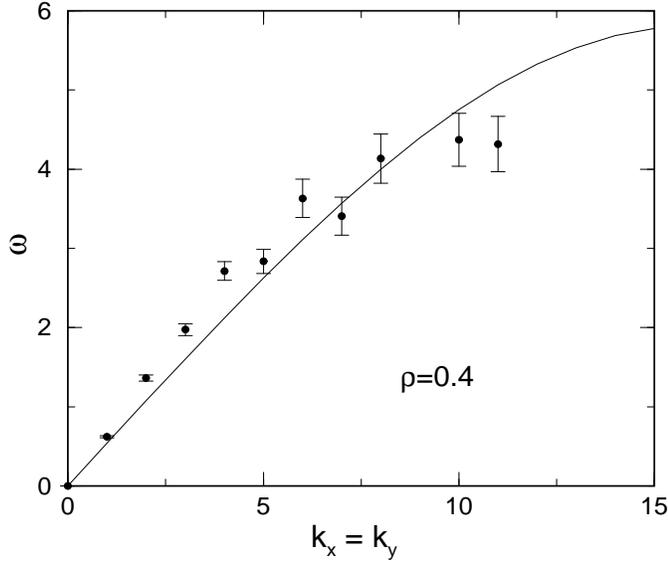,height=3.0in,width=3.5in}
\caption{The excitation spectrum. The solid line is the spin-wave
result. It is linear for small $|{\bf k}|$ with a slope (critical
velocity) $c= 2t\sqrt{1-\bigl ({\mu \over {4t}}\bigr )^2}$.}
\label{disp-L32}
\end{figure}

As shown in the previous section, the superfluid is stable, the low
lying excitations are phononic since the dispersion is linear in
$|{\bf k}|$, Eq.~\ref{phonon}. In Fig.~\ref{disp-L32} we show
$\omega({\bf k})$ for $\rho=0.4$. The solid line is from spin-waves
while the circles are the numerical results. Numerically, $\omega({\bf
k})$ was estimated by fitting an exponential to $\langle a({\bf
k},\tau) a^{\dagger}({\bf k},0) \rangle$ where $\tau$ is the imaginary
time. This quantity is difficult to measure\cite{troyer}, especially
for large momentum vectors, as we see from the noise in the numerical
results. It is possible to get better numerical results for this
quantity by using more elaborate methods, such as the maximum entropy
method.

We see that spin-waves give an excellent description of the
system. The agreement between the numerical simulations and the
spin-wave calculation is extremely good for all the quantities we
considered, except $\omega({\bf k})$, where the difficulty is
numerical. Furthermore, this agreement extends over the entire density
range, $0\leq \rho \leq 1$.

As seen above, the spin-wave results presented here contain momentum
sums which we do numerically. We can, however, find simple empirical
fits which agree with the simulations at least as well as the
spin-wave calculation. The choice of the functional form of the fits
was guided by two considerations: (a) Particle hole symmetry which
suggests dependence on $\rho(1-\rho)$, and (b) the low density
diagrammatic results in the continuum (see next section). So, we found 
the following fits for the boson energy per site:

\begin{eqnarray}
E_{fit} &=& \rho(1-\rho)\Biggl ( {{4\pi \rho(1-\rho)}\over {|{\rm
ln}(A\rho(1-\rho))|}} \nonumber \\
&&\Bigl (1-B {{{\rm ln}|{\rm ln}(A\rho(1-\rho))|}
\over {|{\rm ln}(A\rho(1-\rho))|}}\Bigr ) -4 \Biggr ),
\label{Efit}
\end{eqnarray}
where $A=0.537$ and $B=3.588$ are the fitting parameters. The
condensate density is given by
\begin{equation}
\rho_0^{fit}= \rho(1-\rho) \Bigl (1- {1\over{|{\rm
ln}(0.0335\rho(1-\rho))|}}\Bigr ),
\label{R0fit}
\end{equation}
and superfluid density,
\begin{equation}
\rho_s^{fit}= \rho(1-\rho) \Bigl (1+ {0.11245\over{|{\rm
ln}(\rho(1-\rho))|}}\Bigr ).
\label{Rsfit}
\end{equation}
We can obtain an equally good fit for the superlfuid density using the 
form
\begin{equation}
\rho_s^{fit}=\rho(1-\rho)\Bigl (1+A\rho(1-\rho)\Bigr ),
\label{Rsfit2}
\end{equation}
with $A=0.3356$. To give an idea of the goodness of these fits, we
mention that they agree as well with the numerical results as the
spin-wave results. It is interesting to note that, although in
Fig.~\ref{RsR0vsrho} $\rho_s$ and $\rho_0$ appear to have similar
functional forms, using Eq.~\ref{Rsfit2} gives very poor results for
$\rho_0$. 

We re-iterate that these fits,
Eqs.~(\ref{Efit},\ref{R0fit},\ref{Rsfit},\ref{Rsfit2}) are purely empirical
and are given just as a shortcut to get accurate results.

\section{The continuum}

The excellent agreement we found in the previous section between
simulations and spin-waves suggests a comparison with the ladder
diagram summation results in the continuum.~\cite{schick,hines} The
point is that the continuum results were obtained using approximations
that are valid for very low densities, which is where the lattice
results should approach the continuum. This should give an idea of the
density where the continuum results start to break down. Of course,
this breakdown could be due to the breakdown of the validity of the
ladder diagram calculation, or to the increasing importance of lattice
effects, or both. 

Using diagramatic methods, Schick~\cite{schick} showed that the
condensate density is given by,\cite{comment}

\begin{equation}
{1 \over {2{\bar \rho_0}}} = -{\rm PP} \int_0^{\infty} {{|f(0,x)|^2}
\over {x^2-{\bar \mu}}} {{dx} \over x},
\label{PP}
\end{equation}
where PP stands for principal part, and ${\bar \mu}= 2\mu a^2, {\bar
\rho_0}= \rho_0 a^2/\pi$. The diameter of the hardcore
particles is $a$. In the lattice simulations this corresponds to the
lattice spacing. $f({\bf q^{\prime}},{\bf q})$ is the scattering
amplitude in two dimensions for two particles interacting via the
hardcore potential. Schick~\cite{schick} showed that

\begin{equation}
|f(0,{\bf q})|^2 = {{16}\over {[J_0^2(qa)+Y_0^2(qa)]}},
\label{fq}
\end{equation}
where $J_0^2(qa)$ and $Y_0^2(qa)$ are Bessel functions of zero order
of the first and second kind, respectively. By considering the first
few terms in the small argument expansion of the Bessel functions,
Hines {\it et al} performed the integral Eq.~\ref{PP} and obtained for
the condensate density,\cite{hines}

\begin{eqnarray}
{1\over {2{\bar \rho_0}}}&=& {{-8\pi^2{\rm ln} {\bar \mu}}\over {{\bar
\mu} ({\rm ln}^2{\bar \mu} + \pi^2)}}+ {{16\pi^2(\gamma-{\rm ln}2)({\rm
ln}^2{\bar \mu}-\pi^2)}\over {{\bar \mu} ({\rm
ln}^2{\bar \mu}+\pi^2)^2}}\nonumber \\
&& + {\cal O}\Bigl ({1\over {{\bar \mu}{\rm
ln}^3{\bar \mu}}}\Bigr ).
\label{rho0hines1}
\end{eqnarray}
Where $\gamma$ is Euler's constant. Using this equation with the
appropriate Green functions~\cite{schick}, we can obtain the relation
between the chemical potential and particle density,

\begin{eqnarray}
\mu &=& {{8\pi \rho}\over -{\rm ln}{\bar \rho}}\Biggl (1-{{{\rm
ln}(-{\rm ln}{\bar \rho})}\over {-{\rm ln}{\bar \rho}}} + 
{{(2\gamma +{\rm ln}2 + 2{\rm ln}\pi-1)}\over {-{\rm ln}{\bar
\rho}}}\nonumber \\
&& + {{{\rm ln}^2(-{\rm ln}{\bar \rho})}\over {(-{\rm ln}{\bar
\rho})^2}} - (4\gamma +2{\rm ln}2 + 4{\rm ln}\pi-1){{\rm ln}(-{\rm
ln}{\bar \rho}) \over {(-{\rm ln}{\bar \rho})^2}} \nonumber \\
&&+ {\cal
O}\Bigl ({1\over {{\rm ln}^2{\bar \rho}}} \Bigr )\Biggr ).
\label{muhines}
\end{eqnarray}
Integrating this equation with respect to the density yields the
energy per particle,

\begin{eqnarray}
\epsilon_p &=& {{4\pi \rho}\over -{\rm ln}{\bar \rho}}\Biggl (1-{{{\rm
ln}(-{\rm ln}{\bar \rho})}\over {-{\rm ln}{\bar \rho}}} +
{{(2\gamma +{\rm ln}2 + 2{\rm ln}\pi-3/2)}\over {-{\rm ln}{\bar
\rho}}}\nonumber \\
&& + {{{\rm ln}^2(-{\rm ln}{\bar \rho})}\over {(-{\rm ln}{\bar
\rho})^2}} - (4\gamma +2{\rm ln}2 + 4{\rm ln}\pi-2){{\rm ln}(-{\rm
ln}{\bar \rho}) \over {(-{\rm ln}{\bar \rho})^2}} \nonumber \\ &&+
{\cal O}\Bigl ({1\over {{\rm ln}^2{\bar \rho}}} \Bigr )\Biggr ).
\label{ephines}
\end{eqnarray}
Finally, we can also get the condensate density,

\begin{equation}
\rho_0 = \rho\Bigl (1- {1\over {-{\rm ln}({\bar \rho})}} +{\cal
O}\bigl ({1\over {{\rm ln}^2{\bar \rho}}} \bigr ) \Bigr ).
\label{R0hines}
\end{equation}

To compare these equations with the spin-wave and simulation results,
we should first recall that the kinetic energy term used there is the
full Laplacian. On the other hand, for the lattice bosons we did not
include the diagonal term of the Laplacian in the kinetic
energy. Furthermore, we normalized the energy by the number of
sites. So, the simulation and spin-wave energies should be compared
with $\rho(\epsilon_p-4)$. With this proviso,
Eqs.~(\ref{muhines},\ref{ephines},\ref{R0hines}) are used to calculate
the dashed lines in
Figs.~(\ref{Evsrho-L32},\ref{R0vsrho-L32},\ref{rhovsmu-L32}). We see
that in all cases the ladder calculation agrees well with both the
lattice simulations and the spin-wave calculations for densities
$0\leq \rho\leq 0.05$ and sometimes up to $\rho=0.1$. It is
interesting to note that these densities are mucg higher than was
thought to be the range of validity of the ladder
approximation.\cite{hines}

\begin{figure}
\psfig{file=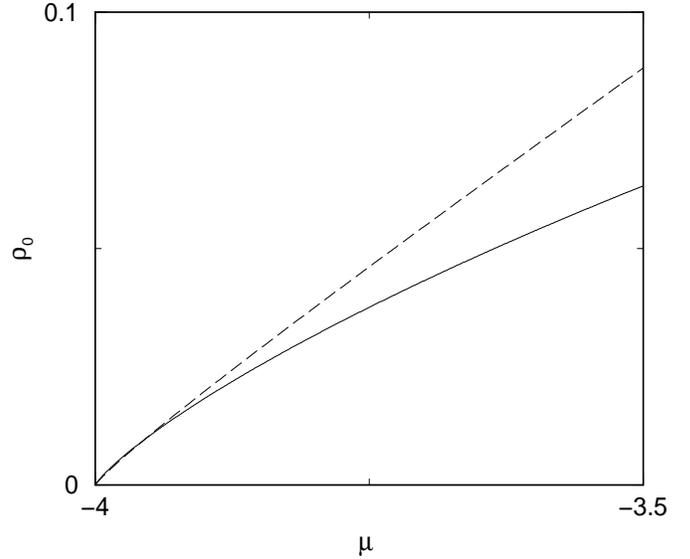,height=3.0in,width=3.5in}
\caption{$\rho_0$ vs $\mu$. The dashed line is the full numerical
integration of Eq.~\ref{PP}, the solid line is the spin-wave result
for a $1024\times 1024$system.}
\label{R0vsmu-SWLADDER}
\end{figure}

One can also integrate Eq.~\ref{PP} numerically thus avoiding the
asymptotic expansion of the Bessel functions. In
figure~\ref{R0vsmu-SWLADDER} we show $\rho_0$ versus $\mu$ from the
numerical integration of Eq.~\ref{PP} (dashed line) and the spin-wave
calculation (solid line). We see that the exact (numerical)
calculation of $\rho_0$ with the ladder approximation does not improve
the comparison with the numerical results. In particular, since this
approximation is based on the assumption of very low density, it does
not exhibit the particle hole symmetry present in the system.

\begin{figure}
\psfig{file=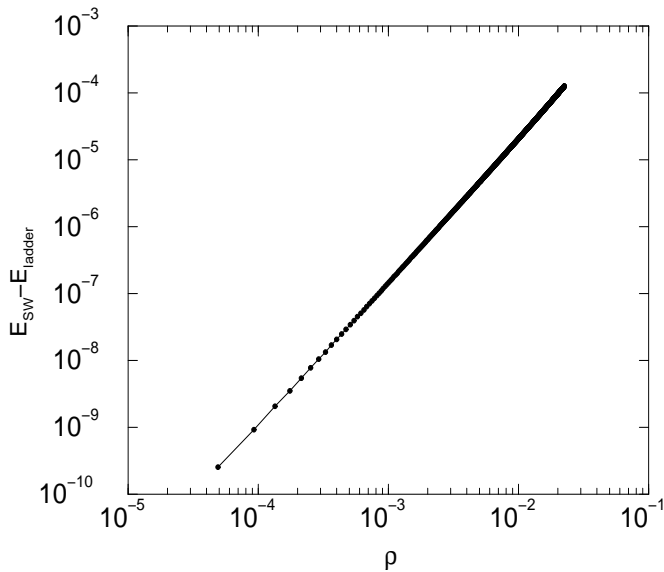,height=3.0in,width=3.5in}
\caption{The difference between the spin-wave and the ladder
diagram energies is plotted as a function of $\rho$. The difference
vanishes like $\rho^2$.}
\label{Esw-Eladdvsrho}
\end{figure}

To demonstrate how agreement between the ladder and spin-wave
approximations is approached, we show in Fig.~\ref{Esw-Eladdvsrho} the
spin-wave energy, Eq.~\ref{ESW}, minus the ladder diagram energy,
Eq.~\ref{ephines}, versus $\rho$. We see that the agreement is indeed
excellent and that the difference tends to zero like $\rho^2$. To get
a clearer idea of the agreement, one should also look at the
fractional difference, $\Delta E/E_{SW}
=(E_{SW}-E_{ladder})/E_{SW}$. For example for $\rho=0.022$ we find
$\Delta E/E_{SW}=1.4\times 10^{-3}$, while for $\rho=5\times 10^{-5}$
we find $\Delta E/E_{SW}=1.3\times 10^{-6}$.

While discussing superfluids in the continuum, we take the opportunity
to make some comments regarding the Onsager-Penrose estimate of the
condensate density.\cite{onsager} Often in the literature (see for
example reference ~\onlinecite{refonsager} and references therein)
results for the condensate density, $\rho_0$, or condensate fraction,
$\rho_0/\rho$, are compared with the estimate by Penrose and
Onsager\cite{onsager} as the standard calculation. However, the
estimate by Penrose and Onsager is incorrect because the trial
wavefunction they use it too restrictive: It is unity if the particles
do not touch and vanishes if they do touch. In fact, wavefunctions do
not vanish so brutally when they approach an infinite wall, they go to
zero gradually: They start decreasing {\it before} the wall is
reached. It turns out that this effect, neglected in the calculation
of Ref.~\onlinecite{onsager}, changes the results a great
deal.\cite{lebellac} In three dimensions, according to
Ref.~\onlinecite{onsager}, $\rho_0\approx \rho(1-(4\pi/3)a^3\rho)$
while using the approach of Bogoliubov\cite{bog1}, one
finds\cite{lebellac} $\rho_0\approx \rho(1-(8/3)\sqrt{(a^3\rho)/\pi}$,
where $a$ is the diameter of the particles. It so happens that for the
physical values present in experiments, both estimates give
$\rho_0/\rho)$ of the order of $10\%$. However, they give very
different functional dependence on the density, $\rho$. Similar
differences occur in two dimensions, of course.

\section{Finite Temperature}

In addition to the ground state, we studied the finite temperature
Kosterlitz-Thouless transition to superfluidity as a function of the
particle density.

\begin{figure}
\psfig{file=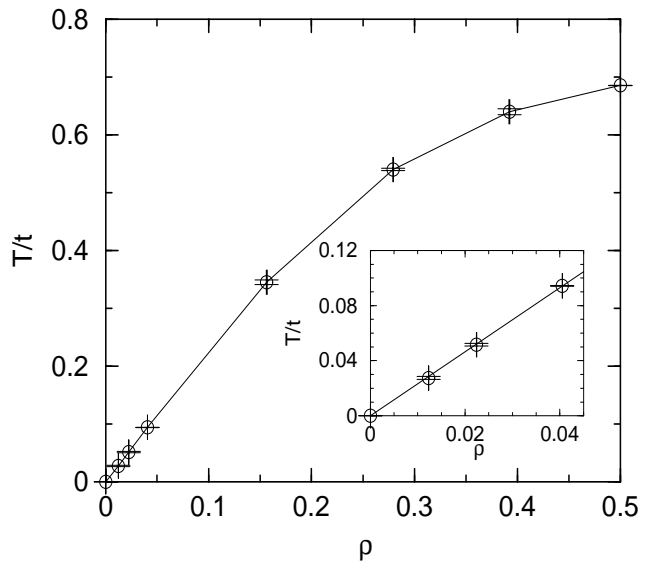,height=3.0in,width=3.3in}
\caption{The critical temperature, $T_{KT}$, as a function of the
particle density. The inset shows $T_{KT}$ for a small density range
$0\leq \rho\leq 0.045$.}
\label{rho_Tc.eps}
\end{figure}

To determine the transition temperature, $T_{KT}$ we exploit the
universal jump in the superfluid densiy shown in Eq.~\ref{rhosKT}. The 
results are show in Fig.~\ref{rho_Tc.eps}. The simulations were done
for $L\times L$ systems with $L=16, 32, 64, 96$ and extrapolated to
the thermodynamic limit. 

For small, but not too small, particle densities, the bosons behave
approximately like free particles which should give $\rho_s \propto
\rho$. The simulations, Fig.~\ref{rho_Tc.eps}, confirm this. We get,
\begin{eqnarray}
\rho_{s}\frac{\pi}{2}\frac{\hbar^2}{m}=k_BT_{KT} 
\simeq 0.740(2) \frac{\pi}{2}\frac{\hbar^2}{m} \rho. 
\label{TKTlin}
\end{eqnarray}
On the other hand, for very low densities, and consequently very low
$T_{KT}$, we should see a crossover to a nonlinear regime when the
correlation length, $\xi$, exceeds the mean interparticle distance,
$\rho^{1/d}$. Using the results of Fisher and
Hohenberg\cite{hohenberg}, the $\rho$ dependence of $T_{KT}$ in this
limit is given by
\begin{eqnarray}
k_BT_{KT}\simeq \frac{4\pi \hbar^2 \rho}{2m \ln \ln (1/(\rho a^2))}
\label{TKThohn}
\end{eqnarray}
where $a$ is the scattering length, equal to one lattice spacing in
the case of hardcore bosons.  Using equations \ref{TKTlin} and
\ref{TKThohn}, we get an estimate for the upper bound of the density
at which the crossover occurs,
\begin{eqnarray}
\rho a^2 < e^{-e^{4/0.740(2)}} < 10^{-90}.
\label{toosmall}
\end{eqnarray}
A value which is far too small to be tested numerically.  This gives
some limit on the usefulness of the result of Fisher and Hohenberg.

\section{Conclusions}

By direct comparison with numerical simulations, we have shown that
spin-waves describe extremely well the properties of hardcore bosons,
or equivalently spin-$1/2$ Heisenberg antiferromagnet in a magnetic
field. This remarkable agreement extends over the entire density
range, $0\leq
\rho \leq 1$.

This verification of the accuracy of the spin-wave approach in the
absence of interactions other than hard-core contact, offers a measure
of confidence in the accuracy of such an analysis even in the presence
of near and next near interactions.\cite{ggb1} We believe a mean field
plus spin-wave approach should still work well in this case. A direct
comparison with numerical simulations would, however, offer more
conclusive assurance. Of course, the quality and reliability of the
spin-wave calculation depends strongly on the mean field states that
are assumed therein lies the challenge. In the case of pure hardcore
repulsion, the mean field state was easy to guess: Uniform particle
density. However, the presence of competing interactions, such as near
(nn) and next near (nnn) neighbors, can complicate things. For
example, if the nn interaction dominates, it seems reasnoable to use a
checkerboard density wave as the mean field state. However, there are
some other exotic states that might lower the energy and need to
examined. One such candidate is the bond-ordered state.\cite{sachdev}

The agreement between the spin-waves and the ladder diagram
approximation for small densities, not only bears out the exact
leading term of the energy, $\epsilon(\rho)=4\pi
\rho (\hbar^2/2m)|{\rm ln}(\rho a^2)|^{-1}$, but also agrees very well
with the logarithmic corrections. For example, Eq.~\ref{ephines}
agrees very well with the numerical results, Fig.~\ref{Evsrho-L32}, up to
$\rho\approx 0.12$, whereas, keeping only the leading contribution,
deviations are observed at $\rho\approx 0.07$. In other words, the
numerical results are accurate enough to probe the logarithmic
corrections. It would therefore be interesting to compare numerical
simulations of hard-core bosons in the continuum with our spin-wave
calculation for large densities. We could not find such simulations in
the literature. One interesting question in that regard is whether the
continuum limit of the spin-wave equations accurately describes bosons
in the continuum. In that case, the spin-wave approach would offer an
alternate simpler way to calculate the continuum case.

Finally, we also determined the KT transition temperature, $T_{KT}$,
as a function of particle density. We found that the theoretically
predicted behavior\cite{hohenberg} appears to be valid at such
exceedingly small values as to preclude numerical and experimental
verification.

\vskip0.5in
\centerline{Acknowledgments}

We acknowledge very useful conversations with F. H\'ebert, M. Le
Bellac and R. T. Scalettar. We also thank ETH-Zurich and HLRS
(Stuttgart) for very generous grants of computer time.

\end{document}